\documentstyle[12pt]{article}

\def\lsim{\stackrel{\textstyle <}{_{\textstyle \sim}}}
\def\gsim{\stackrel{\textstyle >}{_{\textstyle \sim}}}
\def\beqn{\begin{eqnarray}}
\def\eeqn{\end{eqnarray}}
\relax

\relax

\catcode`\@=11

\def\@normalsize{\@setsize\normalsize{15pt}\xiipt\@xiipt
\abovedisplayskip 14pt plus3pt minus3pt%
\belowdisplayskip \abovedisplayskip
\abovedisplayshortskip  \z@ plus3pt%
\belowdisplayshortskip  7pt plus3.5pt minus0pt}

\def\small{\@setsize\small{13.6pt}\xipt\@xipt
\abovedisplayskip 13pt plus3pt minus3pt%
\belowdisplayskip \abovedisplayskip
\abovedisplayshortskip  \z@ plus3pt%
\belowdisplayshortskip  7pt plus3.5pt minus0pt

\def\@listi{\parsep 4.5pt plus 2pt minus 1pt
            \itemsep \parsep
            \topsep 9pt plus 3pt minus 3pt}}

\def\underline#1{\relax\ifmmode\@@underline#1\else
	$\@@underline{\hbox{#1}}$\relax\fi}
\@twosidetrue

\catcode`@=12

\def\FERMICONF{}
\def\FERMILABConf#1{\def\FERMICONF{#1}}
\def\ps@headings{\def\@oddfoot{}\def\@evenfoot{}
\def\@oddhead{\hbox{}\hfill
	\makebox[.5\textwidth]{\raggedright\ignorespaces --\thepage{}--
	\hfill {\rm FERMILAB--Conf--\FERMICONF}}}
\def\@evenhead{\@oddhead}
\def\subsectionmark##1{\markboth{##1}{}}
}

\ps@headings

\relax
\FERMILABConf{92/246--T}
\begin{document}
\begin{titlepage}
\def\ba{\begin{array}}
\def\ea{\end{array}}
\def\thefootnote{\fnsymbol{footnote}}
\begin{flushright}
	FERMILAB--CONF--92/246--T\\
	September 1992
\end{flushright}
\vfill
\begin{center}
{\large \bf ANOMALOUS GAUGE-BOSON COUPLINGS\\
       AT HADRON SUPERCOLLIDERS}\\
\vfill
	{\bf  G.~Valencia}\footnotemark[1]\\
\footnotetext{[*] Talk presented at the XXVI International Conference on
High Energy Physics. Dallas, 1992.}
	{\it Fermi National Accelerator Laboratory \\
	P.O. Box 500, Batavia, IL 60510}
	%\footnote{text}\\[.2in]
\vfill
     %	{\large \bf Abstract}
\end{center}
\begin{abstract}

We discuss anomalous gauge boson couplings at
hadron supercolliders. We review the usual description of these
couplings, as well as the studies of a strongly interacting electroweak
symmetry breaking sector. We present
an effective field theory formulation of the problem
that relates the two subjects, and that allows a consistent and systematic
analysis. We end with some phenomenology.

\end{abstract}
\end{titlepage}

\section{Introduction}

The main reason for building hadron supercolliders like the SSC or the
LHC is to study the mechanism that breaks electroweak symmetry. There
are many possibilities for this, and each one has distinctive signals.
The simplest one is the minimal standard model with a Higgs boson, in
which case the goal would be to find this particle. Typical technicolor
models contain a large number of particles that should be found by these
colliders; the  case of the technirho has been studied most. Similarly,
other possibilities like supersymmetry should have several particles within
reach.

An interesting question is whether any of these new particles will be within
the reach of the SSC/LHC. In anticipation that all new particles could
lie just out of reach, many studies have been undertaken to extract
information on electroweak symmetry breaking from its indirect effects.
There are two fields in the literature that address this issue. One goes
by the name of ``anomalous gauge-boson couplings'', and the other one is
known as ``strong longitudinal gauge-boson scattering''. In this talk
we will present an effective field theory formulation of the problem,
that relates the two subjects and
that allows a consistent and systematic study of anomalous couplings.

\section{Anomalous Couplings}

Conventional studies of anomalous couplings start from the Lagrangian$^1$
\beqn
{\cal L}&=&-ig_V \bigg[ g_1^V(W^\dagger_{\mu\nu}W^\mu-W^{\dagger\mu}W_{\mu\nu})
V^\nu
+ \kappa_V W^+_\mu W^-_\nu V^{\mu\nu}
+{\lambda_V \over M^2_W}W^\dagger_{\lambda\mu}W^\mu_\nu
V^{\nu\lambda}\bigg] \nonumber\\
&+& a_c (W^+_\mu W^-_\mu)^2 + \cdots
\label{ca}
\eeqn
where $V$ is either a photon or a $Z$-boson, and $\cdots$
stands for all other terms that we have not written.

There are several questions that arise when this Lagrangian is used. The
first one is that it is not obviously $SU_L(2)\times U_Y(1)$ gauge
invariant.$^2$
Although this has caused some confusion in the literature, it is {\it not}
a problem, as was recently emphasized by Burgess and
London.$^3$  The reason why
this is not a problem is that this Lagrangian is the unitary gauge version
of an explicitly gauge invariant equivalent Lagrangian. For practical
applications to supercolliders, it turns out to be useful to work with
the explicitly gauge-invariant version. In particular, because this
permits the use of the equivalence theorem to simplify the calculations.

Another issue that is not clear in Eq.~(\ref{ca}), is the question
of how many independent anomalous couplings there are (the $\cdots$), and
if there is a hierarchy amongst them.

The use of Eq.~(\ref{ca}) also creates problems at the one-loop level.
For example, the standard model at one-loop generates some
of the ``anomalous'' couplings and one needs a procedure to separate these
contributions from others due to electroweak symmetry breaking.
Also, the Lagrangian of Eq.~1 is not renormalizable, and no procedure
has been specified to treat the divergences that arise beyond tree-level.
Some authors have addressed this problem at the practical level by
introducing form factors.$^1$ Doing so, further complicates the issue of
counting and classifying the independent couplings.

We answer all these questions by using an effective (``chiral'')
Lagrangian supplemented with the rules of chiral perturbation theory.$^4$ This
gives us an effective field theory formalism that allows a systematic
and consistent study of these issues.

\section{Strongly Interacting $V_L$}

Much work has been done under the heading of strongly interacting
longitudinal $W$'s and $Z$'s ($\equiv V_L$).$^{5,6}$ Perhaps the most important
concept in this field is that of ``enhanced electroweak strength''. In the
standard model, the amplitude for $WW$ scattering is proportional to
$g^2$ multiplied by polarization vectors for the $W$'s. If one looks
at the longitudinal polarization (in a frame where $E/M_W \gg 1$),
one finds that some of the $g^2$ terms are now multiplied by $M_H^2/M_W^2$.
For a very heavy Higgs boson, these terms are thus much larger than the
usual $g^2$ terms, they are of ``enhanced'' electroweak strength.$^{5,6}$
In the
standard model without a Higgs boson the Higgs mass is replaced by the
energy of the $W$'s. One then has terms of order $g^2$, terms
of order $ g^2 s/M_W^2 \sim s/v^2$, $v\approx 250$~GeV, and
also terms of order $(s/v^2)(s/\Lambda^2)^n$ where $\Lambda$ is the scale
at which the effective theory breaks down.
The terms that grow with $s$ are of enhanced
electroweak strength at high energies.
It is these terms that
are of interest at hadron supercolliders. When one
is only interested in extracting these terms, one can resort to the
equivalence theorem and replace the gauge bosons $W,Z$ with their would-be
Goldstone bosons $w,z$ even inside loops.$^{5b,5c,7}$

By computing $V_L$ scattering in the standard model, one is
able to place ``unitarity'' bounds on the Higgs mass.$^{5a,5b}$
At high energies,
the partial waves are proportional to $M_H^2$, and thus if $M_H$ is too
large, they violate the unitarity condition $|a|\leq 1$. In the absence
of the Higgs boson, these partial wave amplitudes grow like $s$, violating
the unitarity bound at about $2$~TeV. One expects the physics associated with
electroweak symmetry breaking to come in at this scale (or below). In the
case where no new particles lie below this scale, the study of anomalous
couplings consists of looking for deviations of the leading $s$
behavior of amplitudes at high energies. One can think of this as a general
way of taking the infinite mass limit of the standard model Higgs.

\section{Effective Lagrangian}

To construct the effective Lagrangian we introduce the Goldstone bosons
$w^\pm$, $z$ through the matrix $U=\exp(i\vec{\tau}\cdot\vec{w}/v)$, and
the gauge fields through the covariant derivative
$$
D_\mu U = \partial_\mu U +i {g^\prime \over 2}B_\mu \tau_3 U -
i {g \over 2} U W_\mu
$$
We also have the field strength tensors
\begin{eqnarray}
W_{\mu\nu}&=&{1 \over 2}\bigg(\partial_\mu W_\nu -\partial_\nu W_\mu
-i {g \over 2} [W_\mu,W_\nu]\bigg)
\nonumber\\
B_{\mu\nu}&=&{1 \over 2}\bigg(\partial_\mu B_\nu -\partial_\nu B_\mu\bigg)
\tau_3
\end{eqnarray}
The lowest order effective Lagrangian is:$^8$
$$
{\cal L}^{(2)}={v^2 \over 4} {\rm Tr}D_\mu U^\dagger D^\mu U +\cdots
$$
where $\cdots$ stands for the usual gauge boson kinetic terms, couplings to
fermions, and gauge fixing terms. This is the term that gives the $W$
and $Z$ their mass as can be seen immediately in unitary gauge ($U=1$).
This is a non-renormalizable Lagrangian, and divergences that occur at
the loop level are handled in the usual way by chiral perturbation
theory. At tree-level, this Lagrangian produces amplitudes of order $E^2$
in an energy expansion. At one-loop, the divergences that appear are of
order $E^4$, and are absorbed by renormalization of the next to leading
${\cal O}(E^4)$ effective Lagrangian.

By looking for the most general form consistent with a global $SU(2)\times
U(1)$
symmetry spontaneously broken to $U(1)$ (but conserving CP), Longhitano
found$^8$ that the next to leading order Lagrangian contains 13 terms
(and the leading order Lagrangian contains an extra term, $\Delta \rho$).
We will
not use this general Lagrangian, but instead we will introduce an additional
assumption: that there is a custodial $SU(2)$ in the physics of electroweak
symmetry breaking. That is, that the global symmetry is $SU(2)\times SU(2)$
broken to $SU(2)$. This amounts to requiring that the additional ``custodial''
$SU(2)$ be broken only by $g^\prime$ and by the difference in fermion masses.

This is a reasonable assumption, which is true both for the minimal standard
model and for common extensions such as technicolor. It is also a consistent
assumption for experiments at supercolliders. The reason is that at the
very high energies of these machines we can concentrate only on those
terms of enhanced electroweak strength as explained before. The counterterms
needed to renormalize the loop diagrams of enhanced electroweak strength
respect this custodial $SU(2)$. Note that this is {\it no longer valid} for low
energy experiments, such as the ones that will be carried out at LEP2. In
that case the distinction between electroweak strength and enhanced
electroweak strength is not meaningful, and for consistency one must keep
the full counterterm structure that appears at one-loop.

The next to leading order effective Lagrangian is:$^9$
\beqn
{\cal L}^{(4)}&=&
{v^2 \over \Lambda^2}\bigg\{ gg^\prime L_{10}{\rm Tr}\bigg(
U^\dagger B_{\mu \nu} U W^{\mu \nu}\biggr) \nonumber\\
&-&ig L_{9L}{\rm Tr}\bigg(W_{\mu\nu}D^\mu U^\dagger
D^\nu U \bigg)
-ig^{\prime} L_{9R}{\rm Tr}\bigg(B_{\mu\nu}D^\mu U
D^\nu U^\dagger  \bigg) \nonumber\\
&+& L_1 \bigg[{\rm Tr}\bigg(D_\mu U^\dagger D^\mu U\bigg)
\bigg]^2
+ L_2 \bigg[{\rm Tr}\bigg(D_\mu U^\dagger D_\nu U\bigg)
\bigg]^2 \bigg\}
\label{la}
\eeqn
The scale $\Lambda$ is determined by the mass of the lightest particle
in the symmetry breaking sector, and in
any case it is $\Lambda \lsim 4 \pi v$.
The next to leading order terms in the effective Lagrangian
have been normalized by $v^2/\Lambda^2$; this reflects the fact that they
appear as the heavy physics associated with the scale $\Lambda$ is integrated
out. With this normalization, the $L_i$ are all expected to be ${\cal O}(1)$.
By going to unitary gauge one can easily see what are the contents
of this Lagrangian. The first line contains a correction to the $Z$
self energy that has been thoroughly discussed:$^{10}$
$$
L_{10}^r(M_Z)=-\pi S
$$
when we take $\Lambda = 4 \pi v$.
The next line contains the lowest order anomalous three gauge boson
couplings.$^{11,12}$ The more common $\kappa_\gamma$, $\kappa_Z$
and $g_1^Z$ are
simply some linear combinations of $L_{9L}$, $L_{9R}$ and $L_{10}$;$^{12}$
$$
g_1^Z-1 \sim
\kappa_V -1 \sim {\cal O} \bigg( g^2 L_{9L,9R,10} {v^2 \over \Lambda^2}\bigg)
$$
Finally, the last line contains the lowest order anomalous four gauge
boson couplings. There are only two of them: $L_1$ and $L_2$.$^{13}$

The bare $L_i$ coupling constants that appear in the Lagrangian are
used to absorb the divergences generated by ${\cal L}^{(2)}$ at one-loop.
It is the renormalized running couplings $L_i^r(\mu)$ that are
physical and can be related to observables. A convenient renormalization
scheme has been defined in the literature,$^9$ and dimensional regularization
is typically used. Once again, since we are only interested in
terms of enhanced electroweak strength, there is no need to renormalize
the electroweak gauge sector. This is the reason why we do not
need custodial $SU(2)$ breaking counterterms like $\Delta \rho$ (or ``T'').

Some of the anomalous couplings in Ref.~1a, namely
those with $\lambda_V$, are not present in the effective Lagrangian at
order $E^4$. These terms appear at the next order,
$E^6$, and are thus
expected to produce much smaller effects
(suppressed by $\sim s/ \Lambda^2$) than the $\kappa_V$ terms.
They are of the same order as the slope terms introduced when
$\kappa_V$ is modified with a form factor. Within our
assumptions, these terms in Eq.~(\ref{ca}), should have been
normalized by $\Lambda^2$ instead of $M_W^2$.
The energy expansion breaks down at some scale
near 2~TeV, where all the terms become equally important.

We have emphasized that the $L_i^r(\mu)$ couplings are naturally
of order one. A value much larger than 1
of one or more of these couplings, would indicate that the formalism
is breaking down at much lower energies than it should. This is associated
with the presence of some new, relatively light, particle beyond the standard
model. In our Lagrangian, we have explicitly included all the known particles
in the standard model and we have assumed that any new particles associated
with electroweak symmetry breaking are heavy: of order a few TeV. The effect
of these heavy particles is only felt indirectly through the anomalous
couplings. If there are some relatively light particles, for example a
300~GeV Higgs boson, then the formalism has to be modified to include the
light Higgs explicitly in the Lagrangian. For this example, there exists
another formulation of the effective Lagrangian that one could use,
namely that in which the symmetry breaking is linearly realized.$^2$
For studies at the SSC it is reasonable
to assume that there are no such light particles, since they would be
discovered directly. For studies at lower energy machines like LEP2 this is
not the case. A 300~GeV Higgs boson would still not be seen directly and
the study of its indirect effects remains interesting. Of course this is
not the only possibility, there could be, for example, a 300~GeV vector
resonance. To study that case at LEP2 one could use an effective Lagrangian
that contains this field explicitly.$^{14}$

\section{Phenomenology}

The explicit gauge invariance of Eq.~(\ref{la}) allows us to use
the equivalence theorem to simplify the calculations.
As long as we are only interested in terms of enhanced
electroweak strength, we can compute with the ${\cal O}(E^4)$
terms presented here,
replacing all the vector bosons with their corresponding would-be
Goldstone bosons. The only exception is for vector bosons
in the initial state since these couple to light fermions.
For $q\bar{q}$ annihilation we must keep the ``initial'' vector boson.
For vector boson scattering, the effective
luminosity of transverse gauge bosons in the protons is much larger than
that of longitudinal gauge bosons.$^{15}$
In practice, we find that for energies
above $\sim 500$~GeV, the longitudinally polarized initial states completely
dominate the cross sections.$^9$

There are three mechanisms to produce vector boson pairs at hadron colliders.
Each of them is sensitive to different anomalous couplings. The largest source
of vector boson pairs is $q\bar{q}$ annihilation.$^{16}$
This process is sensitive to
anomalous three gauge boson couplings $L_{9L}$ and $L_{9R}$ (also to $L_{10}$
but not in terms of enhanced electroweak strength).$^{9,12}$
The vector boson fusion
mechanism is sensitive to all the anomalous couplings, but only to
$L_1$ and $L_2$ at the enhanced electroweak strength level.$^{9,13}$
Finally, gluon
fusion is not sensitive to any of the anomalous couplings we have discussed
(to ${\cal O}(E^4)$),
but it is to anomalous couplings of the top-quark $g_A-1$.$^{17}$
It has been argued
in the literature that the vector boson fusion process can be separated
experimentally from the other two by tagging one forward jet.$^{18}$

For our numerical studies we will take $\Lambda = 4 \pi v$, in accord
with our assumption that there are no new particles below a few TeV.

One of the couplings of the effective Lagrangian has already been measured.
A fit to all data by Altarelli$^{19}$ translates into $L^r_{10}(\mu)=
0.5 \pm 1.6$ at $\mu=1500$~GeV. This one doesn't contribute to the
processes of interest at the SSC (enhanced electroweak strength production
of $V_L$ pairs). The UA2 collaboration has reported:$^{20}$
$-2.2 \leq \kappa_\gamma -1 \leq 2.6$. This translates into
$|L_9| \lsim 900$.
This is expected to improve$^{21}$  by a factor of 2 at the Tevatron.
Similar results are expected from LEP2.$^1$ Within our framework this
means that there will not be any significant bounds on $L_9$
before the SSC/LHC.
There are no present bounds on the anomalous four-gauge boson couplings.

We have done a very crude phenomenological analysis, in which we
{\it assume} that
it is possible to measure the polarization of the vector bosons. We have
computed the contribution of the anomalous couplings to the integrated
cross section for $0.5<M_{VV}<1.0$~TeV, and defined the contribution to
be observable if it induces at least a 50\% change in this integrated
cross section. With this we find that:$^9$

The $W_L Z_L$ channel will be sensitive to $L_{9L}^r(\mu) \lsim -3.5$
or $L_{9L}^r(\mu)\gsim 2.5$. If we assume that it is not possible to
extract the longitudinal polarization, the change in the rate
is always less than a few percent.

The $W_L^+ W_L^-$ channel will be sensitive to a combination of $L_{9L}$
and $L_{9R}$ if $L_9 \lsim -4.0$ or $L_9 \gsim 3.0$. Again,
this is assuming that all backgrounds can be eliminated and polarizations
measured.

The $W^+_L W^+_L$ channel is sensitive to a combination of $L_1$ and $L_2$
if $|L_1| \gsim 1.$ or $|L_2| \gsim 1.$

To understand the possible significance of these numbers it is instructive
to compare with previous studies on anomalous three gauge boson couplings at
the SSC that considered all polarizations. Kane, Vidal and
Yuan$^{22}$ found that the
SSC would be sensitive to $|L_9| \gsim 25$ and Falk, Falk and Simmons$^{12}$
found that the
SSC would be sensitive to $L_{9L} \lsim -16$ or $L_{9L} \gsim 7$ by
looking at the $WZ$ channel. This result is consistent with our estimate.
They also found that the SSC would be sensitive
to $L_{9R} \lsim -119$ or $L_{9R} \gsim 113$ from the $W\gamma$ channel.
In this case the bound is weaker because the final state can have at most
one longitudinal polarization, so the amplitude can only grow
as $\sqrt{s}$. In the $WZ$ channel, the leading term in the amplitude
grows like $s$.

We have argued that if there are no new light particles, the $L_i$ should be
of order one and not significantly larger. This implies that to obtain
meaningful bounds on anomalous three gauge boson couplings at the SSC,
an effort to separate the transverse background is necessary. We have not
studied the feasibility of this separation, and more detailed
phenomenology is clearly needed. On the other hand, the $W^+ W^+$ channel
seems to be a very promising one to place significant bounds on anomalous
four gauge boson couplings. This channel is particularly useful because it
is the one with the lowest backgrounds, as has been emphasized in the
literature.$^{23}$

{\bf Acknowledgements}

This work was done in collaboration with J.~Bagger and S.~Dawson. I
thank C. Burgess, S. Willenbrock and D. Zeppenfeld for useful discussions.


\begin{thebibliography}{99}

\bibitem{26}
a.
 K.~Hagiwara, R.~Peccei, D.~Zeppenfeld and
K.~Hikasa, {\it Nucl. Phys.} {\bf B282} (1987) 253;
D.~Zeppenfeld and S.~Willenbrock, {\it Phys. Rev. }
{\bf D37} (1988) 1775; E.~Yehudai, SLAC-Report 383, Ph.D.~Thesis (1991),
and references therein.
b.
G.~B\'{e}langer and F.~Boudjema, UdeM-LPN-TH 80,82.

\bibitem{r}
A.~De~R\'{u}jula, {\it et. al.}, CERN-TH 6272/91.

\bibitem{b}
M.~Chanowitz, H.~Georgi and M.~Golden, {\it Phys. Rev.} {\bf D36} (1987)
1490; C. Burgess, D. London, McGill-92/04; McGill-92/05.

\bibitem{13}
S.~Weinberg, {\it Physica} {\bf 96A} (1979) 327;
J.~Gasser and H.~Leutwyler, {\it Ann. Phys.}
{\bf 158} (1984) 142.

\bibitem{2}
a.
D.~Dicus and V.~Mathur, {\it Phys. Rev.} {\bf D7}
(1973) 3111;
M.~Veltman, {\it Acta Phys. Pol.} {\bf B8}
(1977) 475.
b.
 B.~W.~Lee, C.~Quigg, and H.~B.~Thacker,
{\it Phys. Rev.} {\bf D16} (1977) 1519.
c.
M.~S.~Chanowitz and M.~K.~Gaillard,
{\it Nucl.Phys.} {\bf B261} (1985) 379.

\bibitem{19}
S.~Dawson and S.~Willenbrock,
{\it Phys. Rev.} {\bf D40} (1989) 2880;
M.~Veltman and F.~Yndurain,
{\it Nucl. Phys.} {\bf B325} (1989) 1.

\bibitem{18}
J.~M.~Cornwall, D.~N.~Levin, and G.~Tiktopoulos,
{\it Phys. Rev. } {\bf D10} (1974)1145; {\bf 11} (1975) 972
E.

\bibitem{11}
T.~Appelquist and C.~Bernard, {\it Phys. Rev.}
{\bf D22} (1980) 200.
A.~Longhitano, {\it Nucl. Phys.} {\bf B188} (1981)
118.

\bibitem{us}
J.~Bagger, S.~Dawson and G.~Valencia, Fermilab-Pub-92/75-T.

\bibitem{23}
M.~Peskin and T.~Takeuchi, {\it Phys. Rev. Lett.} {\bf 65}
(1990) 964.

\bibitem{28}
C. Ahn {\it et. al.}, {\it Nucl. Phys.} {\bf B309} (1988) 221;
B.~Holdom, {\it Phys. Lett.} {\bf 258B} (1991) 156.

\bibitem{17}
A.~Falk, M.~Luke, and E.~Simmons,
{\it Nucl. Phys.} {\bf B365 } (1991) 523.

\bibitem{15}
A.~Dobado and M.~Herrero, {\it Phys. Lett.}
{\bf 228B} (1989) 495;
J.~Donoghue and C.~Ramirez, {\it Phys. Lett.}
{\bf 234B} (1990) 361;
S.~Dawson and G.~Valencia, {\it Nucl. Phys.}
{\bf B352} (1991) 27.

\bibitem{bess}
For example see R.~Casalbuoni {\it et. al.}, UGVA-DPT-1992-07-778.

\bibitem{20}
S.~Dawson, {\it Nucl. Phys.}
{\bf B249} (1985) 42.

\bibitem{41}
M.~Duncan, G.~Kane and W.~Repko, {\it Nucl. Phys.}
{\bf B272} (1986) 833.

\bibitem{49}
R.~Peccei and X.~Zhang, {\it Nucl. Phys.} {\bf
B337} (1990) 269.

\bibitem{jt}
V.~Barger {\it et. al.},
{\it Phys. Rev.} {\bf D44} (1991) 1426.

\bibitem{25}
G.~Altarelli, CERN-TH 6525/92.

\bibitem{30}
J.~Alitti {\it et. al.},
{\it Phys. Lett.}
{\bf 277B} (1992) 194.

\bibitem{31}
U.~Baur and E.~Berger, {\it Phys. Rev.} {\bf D41}
(1990) 1476.

\bibitem{29}
G.~Kane, J.~Vidal and
C.-P.~Yuan, {\it Phys. Rev.} {\bf D39} (1989) 2617.

\bibitem{44}
M.~Berger and M.~Chanowitz, {\it Phys.
Lett.} {\bf 263B} (1991) 509; {\bf 267B} (1991)
416.

\end{thebibliography}
\end{document}